\documentclass[conference]{IEEEtran}
\IEEEoverridecommandlockouts
\usepackage{cite}
\usepackage{amsmath,amssymb,amsfonts}
\usepackage{amsmath}
\usepackage{cuted}
\usepackage{breqn}
\usepackage{graphicx}
\usepackage{textcomp}
\usepackage{xcolor}
\usepackage{mathtools}
\usepackage[linesnumbered,algoruled]{algorithm2e}
\usepackage{algpseudocode}
\usepackage{amsthm}
\usepackage{comment}
\usepackage{bm}

\usepackage[all]{background}

\usepackage{stackengine}
\setstackEOL{\\}
\setstackgap{L}{\normalbaselineskip}
\SetBgContents{\color{gray}{\tiny \Longstack{PREPRINT - accepted by IEEE Latin-American Test Symposium (LATS) 2022}}}
\SetBgPosition{2.3cm,1cm}
\SetBgOpacity{1.0}
\SetBgAngle{0}
\SetBgScale{1.8}

\def\BibTeX{{\rm B\kern-.05em{\sc i\kern-.025em b}\kern-.08em
    T\kern-.1667em\lower.7ex\hbox{E}\kern-.125emX}}
\begin{document}
\bstctlcite{IEEEexample:BSTcontrol}
\title{A Temperature Independent Readout Circuit for ISFET-Based Sensor Applications\vspace{-5mm}}
\author{\IEEEauthorblockN{Elmira Moussavi\IEEEauthorrefmark{1},
Dominik Sisejkovic\IEEEauthorrefmark{1},
Animesh Singh\IEEEauthorrefmark{2},
Daniyar Kizatov\IEEEauthorrefmark{1},\\
Rainer Leupers\IEEEauthorrefmark{1}, 
Sven Ingebrandt\IEEEauthorrefmark{2},
Vivek Pachauri\IEEEauthorrefmark{2}, and
Farhad Merchant\IEEEauthorrefmark{1}}
		\IEEEauthorblockA{ \IEEEauthorrefmark{1}Institute for Communication Technologies and Embedded Systems, \IEEEauthorrefmark{2}Institute of Materials in Electrical Engineering 1}
        \IEEEauthorblockA{RWTH Aachen University, Germany}
        \{moussavi, sisejkovic, kizatov, leupers, merchantf\}@ice.rwth-aachen.de, \{singh, ingebrandt, pachauri\}@iwe1.rwth-aachen.de \vspace{-5mm}}
        
\maketitle

\begin{abstract}
The ion-sensitive field-effect transistor (ISFET) is an emerging technology that has received much attention in numerous research areas, including biochemistry, medicine, and security applications. However, compared to other types of sensors, the complexity of ISFETs make it more challenging to achieve a sensitive, fast and repeatable response. Therefore, various readout circuits have been developed to improve the performance of ISFETs, especially to eliminate the temperature effect. This paper presents a new approach for a temperature-independent readout circuit that uses the threshold voltage differences of an ISFET-MOSFET pair. The Linear Technology Simulation Program with Integrated Circuit Emphasis (LTspice) is used to analyze the ISFET performance based on the proposed readout circuit characteristics. A macro-model is used to model ISFET behavior, including the first-level Spice model for the MOSFET part and Verilog-A to model the surface potential, reference electrode, and electrolyte of the ISFET to determine the relationships between variables. In this way, the behavior of the ISFET is monitored by the output voltage of the readout circuit based on a change in the electrolyte’s hydrogen potential (pH), determined by the simulation. The proposed readout circuit has a temperature coefficient of 11.9$ppm/ ^{\circ }C$ for a temperature range of 0-100$^{\circ }C$ and pH between 1 and 13. The proposed ISFET readout circuit outperforms other designs in terms of simplicity and not requiring an additional sensor.
\end{abstract}


\section{Introduction}
P. Bergveld proposed the concept of an ion-sensitive solid-state device in 1970 by combining the principle of metal-oxide-semiconductor (MOS) transistor with a glass electrode to benefit from sensitivity up to a single molecule or ion~\cite{bergveld1970development}. Although the use of solid-state technologies can increase the efficiency and controllability of the system, the temperature can lead to inaccuracies in the measurement data~\cite{abdulwahab2018cmos}. Many readout circuits have been developed to minimize the effects of temperature sensitivity and improve the sensor’s accuracy and sensitivity~\cite{gaddour2020temperature}~\cite{rai2021vertical}. Some reported temperature compensation methods are highly complex and incompatible with portable sensors. This led to proposing a simple and efficient approach to compensate for the temperature effect on the readout circuit. By exploiting the CMOS compatibility of ISFET devices, the temperature effect can be canceled out by using the difference between the two threshold voltages in the ISFET-MOSFET pair, in which both have the same temperature coefficient.

\begin{figure}[t]
\centering
\includegraphics[width=0.55\columnwidth]{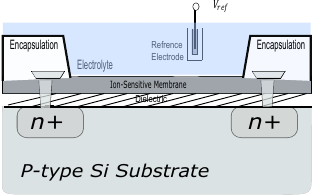}
\caption{Cross section of ISFET sensor structure.}\label{fig:one}
\bigbreak
\includegraphics[width=0.55\columnwidth]{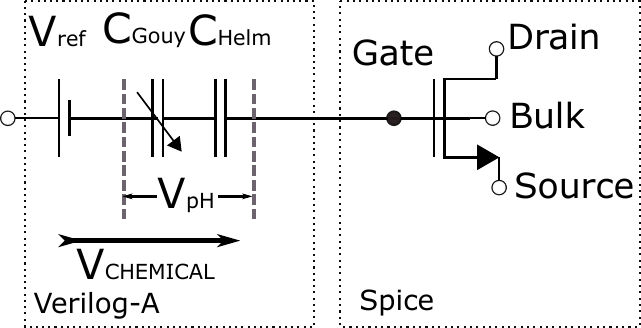}
\caption{ISFET's macro-model circuit emulation.}\label{fig:two}
\vspace{-5mm}
\end{figure}

The ISFET sensor has the same operation as the conventional MOS field-effect transistor (MOSFET), except that the gate-metal layer is created by replacing the MOSFET gate with a special membrane sensitive to the electrolyte solution. The ISFET sensor is exposed to ion diffusion to a sensitive membrane material resulting in a threshold voltage shift~\cite{moussavi2022phgen}. A reference electrode is immersed in the electrolyte as shown in Fig. \ref{fig:one}. The principle of ISFET sensitivity is based on charge absorption at the ion-solid interface between the sensor layer containing the hydroxyl and electrolyte groups~\cite{morgenshtein2003design}. 
ISFET sensors based on field-effect transistors benefit from technological development, especially in microelectronics, allowing miniaturization and mass production, resulting in low-cost components and requiring no special packaging. Nevertheless, low-temperature sensitivity is still required for better measurement accuracy~\cite{harrak2020design}.
One of the most popular ISFET readout circuits is the constant voltage constant current (CVCC) circuit. In the CVCC readout topology, the output signal is monitored by keeping the voltage and current at the drain-source terminals of the ISFET constant. However, the operation is affected by temperature variations and requires an additional temperature sensor combined with a differential operational amplifier as a summation circuit to compensate for the temperature effect. Therefore, we focused on a simple, temperature-independent ISFET readout circuit that does not require additional circuitry for temperature compensation.
 \cite{abdul2009low}\cite{jarmin2014simulation}.
The main contributions are as follows:
\begin{itemize}
  \item The design of a simple readout circuit to eliminate the temperature effect and increase the accuracy of the ISFET output voltage corresponding to the pH value.
  \item The implementation of the readout circuit in such a way that no additional temperature sensor is required. Taking advantage of the compatibility between ISFET devices and conventional CMOS integration circuits, the subtraction of two threshold voltages equally proportional to the absolute temperature is used.
  \item An approach to make the system performance of ISFETs less sensitive to temperature effects over a wide pH range.
\end{itemize}

The rest of the paper is organized as follows. Section II included the ISFET model and theory, as well as the related work on the temperature-independent readout circuit. Section III describes the proposed readout circuit to eliminate the temperature effect. The implementation results are given in Section IV. Finally, a conclusion is drawn in Section V.
\section{Background and Related Works}
\subsection{ISFET Structure and Modeling}
The structure of the ISFET device is based on the conventional MOSFET device. As illustrated in Fig.~\ref{fig:one}, the ISFET sensor is immersed in an electrolyte solution, which creates an electrical potential between the solution and the ion-sensitive membrane described by the Gouy-Chapman model~\cite{FERNANDES2012163}. Besides the semiconductor part, other parts, including the reference electrode and the electrolyte interface, can also extend the temperature effect of the ISFET sensor. To create an accurate model and describe the behavior of ISFETs, a MOSFET model combined with the Helmholtz and Gouy–Chapman model is used. This basic approach provides an efficient model for implementing the readout circuit based on the ISFET macro-model~\cite{Martinoia2000ABM}.
As shown in Fig.~\ref{fig:two}, a Spice model for the MOSFET section is used to describe the behavior of the proposed ISFET readout circuit. Reference electrode potentials ($V_{ref}$) and the double capacitance of Helmholtz-Gouy ($C_{Gouy}, C_{Helm}$), is expected to be modeled effectively to show the temperature effect on the sensor. The simplified model of the ISFET is realized by summing the reference electrode potential and the surface potential $\phi_{0}$ to analyze the flat-band voltage shift ($V_{pH}$). The flat-band voltage shift is linearly proportional to the electrolyte's pH value (hydrogen potential) as follow:
\begin{equation}
V_{pH}= E_{ref}{\frac{Ag}{AgCl}}+\frac{\mathrm{d} E_{ref} }{\mathrm{d} T}(T-298.16)+\chi ^{sol}+\Delta \varphi^{lj}+\Psi_{0}, 
\end{equation}
with
\begin{equation}
\Psi_{0} = -2.303\frac{KT}{q}(pH_{pzc}-pH),
\label{eq.02}
\end{equation}
\begin{figure}[t]
   \centering
   \includegraphics*[width=0.6\columnwidth]{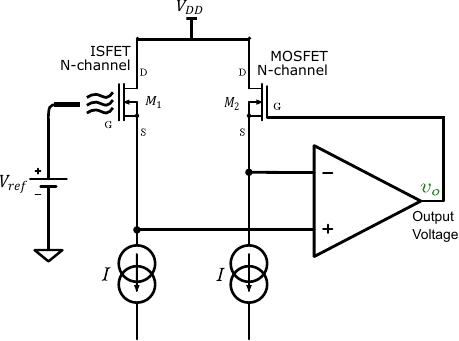}
   \caption{The ISFET-MOSFET pair structure mounted as part of proposed readout circuit.}
   \label{fig:readout}
\end{figure}
\begin{figure}[t]
   \centering
   \includegraphics*[width=0.4\columnwidth]{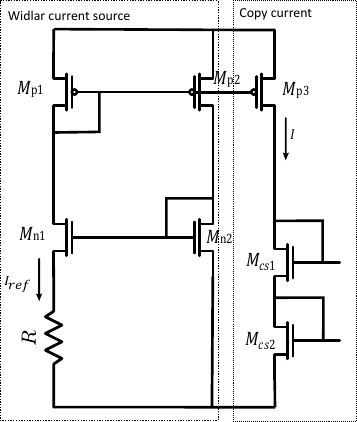}
   \caption{The Widlar current source with current mirror to generate the current for readout circuit's current sources.}
   \label{fig:currentsource}
   \vspace{-5mm}
\end{figure}
where $\frac{KT}{q}$ is the thermal potential, and $pH_{pzc}$ is the pH of point zero charge, T is the system temperature, K the constant of Boltzmann, q the charge, Eref (Ag/AgCl)
the potential of the reference electrode is independent of the temperature, $\Delta\varphi^{lj}$ the potential drop between the solution and the reference electrode, $\chi ^{sol}$ the dipole potential of the surface electrolyte-insulator.
The described model in Eq.~\ref{eq.02} is the simplified one. The more complex model was implemented in Verilog-A for the ISFET simulation model, which consists of the capacitance. The ISFET Spice simulation model is based on the macro-model described in \cite{Martinoia2000ABM} for predicting the behavior of the output signal with respect to the pH value of the electrolyte. 
\subsection{Related Work}
In recent years, work has been reported on compensating for the temperature effect of the ISFET readout circuit using artificial neural networks~\cite{harrak2020design}\cite{bhardwaj2017temperature}. Lower temperature dependence is achieved by training the neural network based on data obtained from the device's interface circuits in different pH solutions at different temperatures. In the model proposed in~\cite{bhardwaj2017temperature}, only pH values above 4 are temperature-independent. However, the model in~\cite{harrak2020design} could obtain a temperature sensitivity $1.5 \times 10^{-4} pH/^{\circ}C$ for the pH range between 1 and 12 by performing an automated control technique to fix the isothermal point. However, the model requires an initial data set from regular samples to predict temporal variations, and and acquiring such a set takes multiple days. In this work, we overcome the temperature effect by taking advantage of the CMOS compatibility of the ISFET. Implementing an ISFET-MOSFET pair with equal temperature coefficients, and using a negative feedback operational amplifier to cancel the temperature effect. Here, no additional complicated circuitry is required.

\section{ISFET Readout Circuit Design with Temperature Effect Compensation}

\begin{figure}[t]
   \centering
   \includegraphics*[width=0.5\columnwidth]{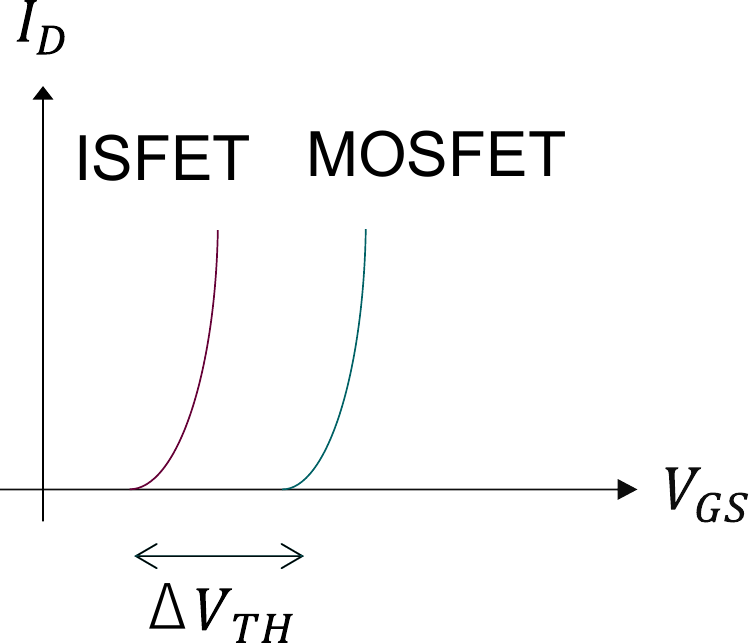}
   \caption{The IV characteristic of ISFET and MOSFET.}
   \label{fig:delta}
   \vspace{-5mm}
\end{figure}

This section proposes a new approach in the readout circuit design to reduce the effects of temperature sensitivity and consequently increase the sensor’s measurement accuracy and controllability.
The schematic of the proposed readout circuit is presented in Fig.~\ref{fig:readout}. A standard CMOS process is used to integrate the ISFET with a readout circuit to improve the readout performance. This circuit allows the threshold voltage extraction in the form of a differential output voltage. The readout circuit consists of the N-channel ISFET-MOSFET pair, where the ISFET transistor ($M_1$) and the NMOS transistor ($M_2$) must be identical, with the same width and length $({W_1}/{L_1}={W_2}/{L_2})$.
The current sources are used to pass the same current into the branches. Thus, a constant current (I) flows through transistors $M_1$ and $M_2$. The Widlar current source is used to fed the current into the circuit via a current mirror (Fig. \ref{fig:currentsource}) to copy the current via transistors $M_{cs1}$ and $M_{cs2}$. The current reference can be generated from the voltage reference by using resistor R. Thus, the current mirror generates a reference current and shares the same current with all blocks of the circuit via current mirrors. 
An operational amplifier is connected to the sources of transistors $M_1$ and $M_2$ to extract the voltage variation caused by the charge concentration in the electrolyte and to compensate for the temperature effect. The voltage difference ($ V_O $) at the output of the op-amp is returned to the gate of transistor $ M_2 $ by a negative feedback. As the transistor $M_1$ is on, the non-inverting input of the amplifier is 
\begin{equation}
   V_{OpampNI}= -V_{GS1},
   \label{eq4}
\end{equation}
 and the inverting input is calculated as
 \begin{equation}
   V_{OpampI}= V_{O}-V_{GS2}.
   \label{eq5}
\end{equation}
The op-amp forces the input terminals (Eq.~\ref{eq4},~\ref{eq5}) to be equivalent; therefore, the $V_O$ can be expressed as follows:
 \begin{equation}
  V_{O}= V_{GS2}-V_{GS1}.
   \label{eq6}
\end{equation}
If all transistors are biased to run in the saturation regime, the following equation can be derived:
 \begin{equation}
  V_{GS}= \sqrt{\frac{2I_D}{{K}'(W/L)}}+V_{TH}.
   \label{eq7}
\end{equation}
Here, $I_D$ is the drain current of the transistor, $W$ is the channel width, $L$ is the channel length, $V_{TH}$ is the threshold voltage, and the technology coefficient is defined as
\begin{equation}
    {K}'=\mu _{n}C_{ox}.
    \label{eq.8}
\end{equation}
The $\mu _{n}$ represents the average carrier mobility on the device channel and $C_{ox}$ is the gate oxide capacitance per unity area. 
A simplified model for ISFET's threshold voltage is given as:
\begin{equation}
V_{TH}= (2.303\frac{KT}{q}\alpha )pH+E_{0},
\end{equation}
where ($\alpha$) is the adjusting parameter and can be between 0 and 1, the second term is independent of pH and temperature, while the first term is directly dependent on the temperature change. To achieve the maximum sensitivity of the ISFET, $\alpha$ is set to 1.
By placing the $V_{GS}$ as described in equation \ref{eq6} to \ref{eq7}, $V_O$ becomes:
\begin{equation}
V_{O}= \sqrt{\frac{2I_{DM2}}{{K}'_2(\frac{W}{L})_2}}+V_{TH_{NMOS}}-\sqrt{\frac{2I_{DM1}}{{K}'_1(\frac{W}{L})_1}}-V_{TH_{ISFET}}.
   \label{eq8}
\end{equation}
Since ISFET devices are essentially MOSFET devices with a modified gate surface, the difference between two N-channel devices (ISFET-MOSFET pair) with different voltage thresholds but proportional to temperature can be used to compensate for the temperature effect.
Where $I_{DM1}= I_{DM2} = I$ and $(W/L)_1=(W/L)_2$ with the same ${K}'$ (Eq.~\ref{eq.8}). Consequently,
\begin{equation}
V_{O}= V_{TH_{NMOS}}-V_{TH_{ISFET}},
   \label{eq9}
\end{equation}
when the voltage $V_{TH_{NMOS}}$ changes due to temperature variation, the voltage $V_{TH_{ISFET}}$ also changes by the same amount. Therefore, the difference ($\Delta V_{TH}$) remains constant and does not vary due to temperature fluctuations. In other words, the temperature coefficient of the threshold voltage is negative, and when the temperature increases, both the N-channel ISFET and the MOSFET change with the same temperature coefficient. Therefore, their difference remains constant, as depicted in Fig. \ref{fig:delta}. Thus, we estimated the condition to have
\begin{equation}
 \frac{\mathrm{d} V_{TH_{ISFET}}}{\mathrm{d} T}=\frac{\mathrm{d} V_{TH_{NMOS}}}{\mathrm{d} T},  
\end{equation}
and consequently:
\begin{equation}
 \frac{\mathrm{d} V_O}{\mathrm{d} T}=0.
\end{equation}
As a result, the differential amplifier circuit is used to eliminate the effect of temperature due to the proportional dependence of voltage on temperature.
We have assumed that the $\mu _{n}$ is the same for ISFET and NMOS, but in reality, they may be slightly different. Nevertheless, they can still achieve reasonable temperature compensation. For the sake of simplicity, channel length modulation is not considered in Eq.~\ref{eq7}. However, the effect is taken into account in the simulation results.

\section{Experimental Results}
\begin{figure}[t]
   \centering
   \includegraphics*[width=\columnwidth]{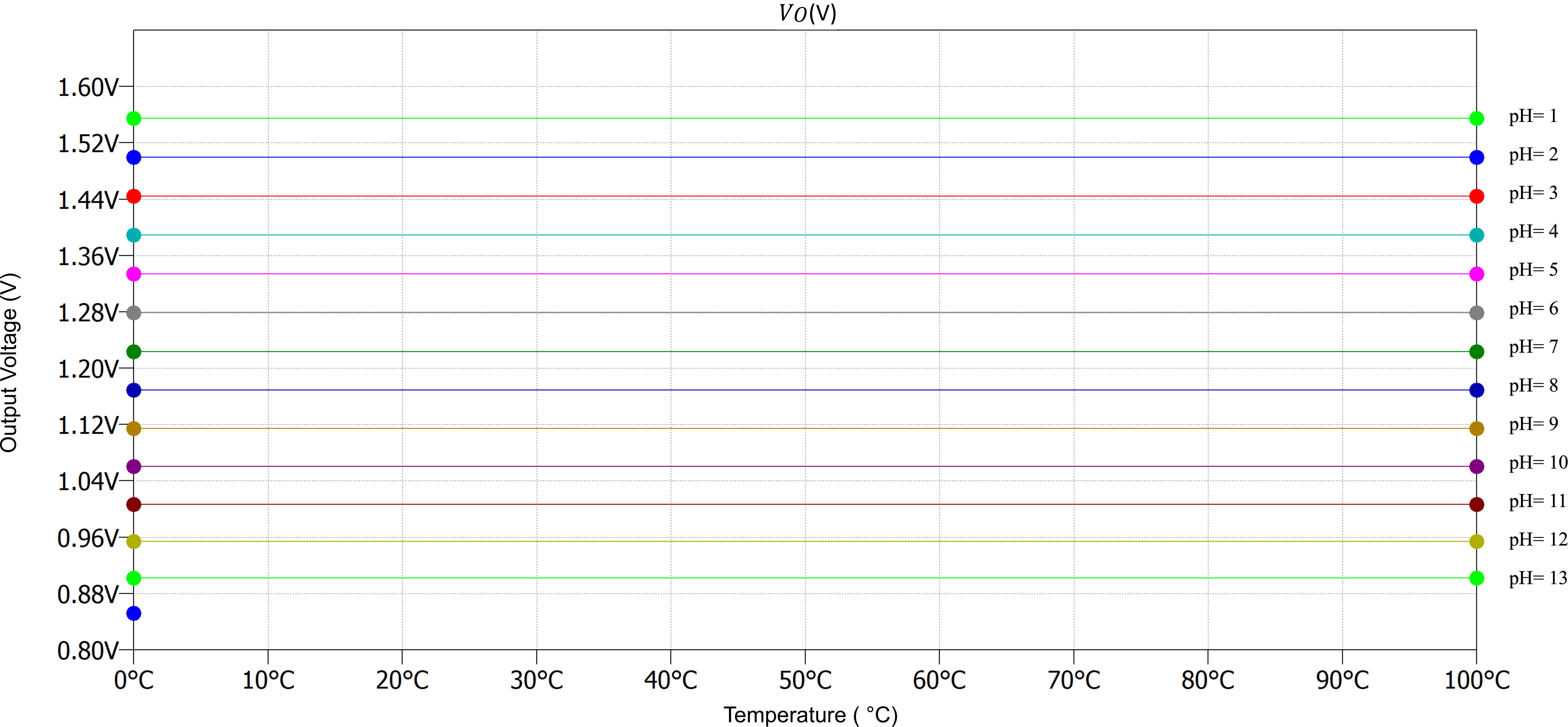}
   \caption{The simulation results of the temperature effect on the output voltage of the proposed readout circuit over pH range of 1 to 13, with $55mV/pH$ and the temperature coefficient of $11.9ppm/ ^{\circ }C$.}
   \label{fig:results}
   \vspace{-5mm}
\end{figure}

The purpose of this readout circuit is to compensate for temperature sensitivity and provide a linear output signal to achieve good measurement accuracy at an exact pH point. The proposed circuit is implemented in LTSpice. The ISFET's macro-model has been extended for the temperature dependence simulation. The ISFET and NMOS transistors have the same dimensions with a length of 18 $\mu m$ and a width of 840 $\mu m$. Simulation is performed to confirm the elimination of the temperature effect of the proposed readout circuit. DC simulation is performed to find the appropriate biasing point for the reference electrode. Furthermore, the pH value varies between 1 to 13 with a temperature change from 0 to 100 $ ^{\circ }C$. The results of the readout simulation show that the temperature coefficient is $11.9ppm/  ^{\circ }C$. The temperature coefficient is calculated as follows:
\begin{equation}
TC(ppm/^{\circ}C)=\frac{V_{O_{max}}-V_{O_{min}}}{V_{O_{mean}}(T_{max}-T_{min})}10^6,
   \label{eq11}
\end{equation}
where $V_{O_{max}}$ and $V_{O_{min}}$ are the maximum and the minimum values of output voltage in the considered temperature (T) range respectively, and $V_{O_{mean}}$ is the mean value.
As illustrated in Fig. \ref{fig:results}, the readout circuit reduces the temperature coefficient. In the simulation for the readout circuit implementation, the ISFET macro-model is used to describe the behavior of the reference electrode, electrolyte, and sensitive membrane. While for the MOSFET part, a standard CMOS model in the Spice model was used. The $18 \mu m$ technology was utilized for the proposed circuit. Consequently, the simulation of the implemented readout circuit shows in Fig.~\ref{fig:results} a temperature coefficient of $11.9ppm/ ^{\circ }C$ for the temperature of $ 0- 100^{\circ}C$, and a pH range between 1 and 13. The main advantage of the proposed readout circuit compared to other reported readout circuits~\cite{harrak2020design}~\cite{abdul2009low}~\cite{bhardwaj2017temperature} with temperature compensation is \emph{the simplicity of the design and the fact that no additional temperature sensor is required}. Also, it is possible to use it for a wide range of pH values between 1 and 13 without requiring initial data to perform the compensation.
The proposed readout circuit is a good candidate for pH sensors with a low temperature coefficient. 
This simulation has shown that it is possible to use the differential threshold voltage of an ISFET-MOSFET pair with the same temperature response to eliminate the temperature effect. For the given operation, the output voltage response per pH is $55mV/pH$ for the ISFET sensor over temperature variations between $ 0- 100^{\circ}C$ by using the proposed readout circuit, which is close to the Nernst limit. The Nernst model defines the maximum achievable voltage sensitivity for pH measurement as $59mV/pH$ at $25^{\circ}C$ for the classical ISFET devices \cite{spijkman2011beyond}.

\section{Conclusion}
Due to higher system efficiency and better controllability, ISFET sensors are used in many applications, especially in biological and medical devices. However, erroneous measurements can be observed due to the influence of temperature variation. This work presents a simple approach to designing an ISFET readout circuit that eliminates the temperature effect for a wide pH range between 1 and 13 with a pH sensitivity of $55mV/pH$ close to the Nernst limit.
Since the threshold voltage of MOSFET and ISFET has a negative temperature coefficient, their threshold voltage shift enables the provision of a temperature-independent output voltage with temperature coefficient of $11.9ppm/ ^{\circ }C$. The readout circuit can be applied to other sensors modified to be sensitive to different types of ions, such as ISFET and ChemFETs.

\section*{Acknowledgement}
This work was partially funded by Deutsche Forschungsgemeinschaft (DFG – German Research Foundation) under the priority programme SPP 2253.

\bibliographystyle{IEEEtran}
\bibliography{bibliography}
\end{document}